\documentclass[twocolumn,showpacs,preprintnumbers,amsmath,amssymb]{revtex4}
\usepackage{graphicx}
\usepackage{dcolumn}
\usepackage{bm}
\begin{document}
\title{Shashkin~{\it et al}.\ reply to cond-mat/0410409}
\author{A.~A. Shashkin$^{1,2}$, S. Anissimova$^1$, M.~R. Sakr$^{1,3}$, S.~V. Kravchenko$^1$, V.~T. Dolgopolov$^2$, and T.~M. Klapwijk$^4$}
\affiliation{$^1$Physics Department, Northeastern University, Boston, Massachusetts 02115, U.S.A.\\
$^2$Institute of Solid State Physics, Chernogolovka, Moscow District 142432, Russia\\
$^3$Department of Physics and Astronomy, UCLA, Los Angeles, California 90095, U.S.A.\\
$^4$Kavli Institute of Nanoscience, Delft University of Technology, 2628 CJ Delft, The Netherlands}
\begin{abstract}
We show that the Comment by Reznikov and Sivan (cond-mat/0410409) is
erroneous because the authors do not distinguish between Pauli and
Curie spin susceptibility.
\end{abstract}
\maketitle

The low-temperature behavior of the dilute electron system in silicon
depends on a delicate interplay between kinetic energy,
electron-electron interaction energy, and disorder. Any expansion of
the strong localization regime to higher electron densities leads to
a fading of the region in which the most pronounced many-body
phenomena are observed. The spin susceptibility in silicon samples
with different levels of disorder tends to diverge at a
sample-independent critical density
$n_\chi\approx8\times10^{10}$~cm$^{-2}$, as was established in
transport \cite{kravchenko02,vitkalov02} and thermodynamic
\cite{shashkin04} measurements. Rather than the absolute values of the
electron density, it is the deviations from this critical point $n_\chi$
that matter. Due to the high level of disorder in their sample, Prus
{\it et al}.~\cite{prus03} could only reach electron densities that
are more than a factor of 2 (rather than by 20\% mentioned in the Comment~\cite{reznikov04}) farther from $n_\chi$ as compared to our experiment. Of course, they could not see any critical many-body phenomena in their sample.

In our paper \cite{shashkin04}, we have studied the clean metallic
regime characterized by the absence of a band tail of localized
electrons. In contrast, Prus {\it et al}.\ have studied mainly the
insulating regime in a highly-disordered sample, in which the band
tail of localized electrons is present at all electron densities
\cite{prus03}. As a result, they have found Curie contribution to the
measured magnetization, which is strongly nonlinear with a magnetic
field, and the extracted spin susceptibility has a Curie temperature
dependence. This is the case even at high electron densities, where
metallic behavior might be expected instead. Such effects are absent
in our samples: the spin susceptibility (in the partially-polarized
system) has been found to be independent of the magnetic field and
temperature. Therefore, there is no overlap between their data and
ours (at least, in the crucial region of low electron densities) as
they were taken in two opposite regimes, contrary to the claim made
by the authors of the Comment \cite{reznikov04}. Apparently, Reznikov
and Sivan do not distinguish between the Pauli spin susceptibility of
band electrons and the Curie spin susceptibility of local moments.

\begin{figure}[b]
\scalebox{0.45}{\includegraphics{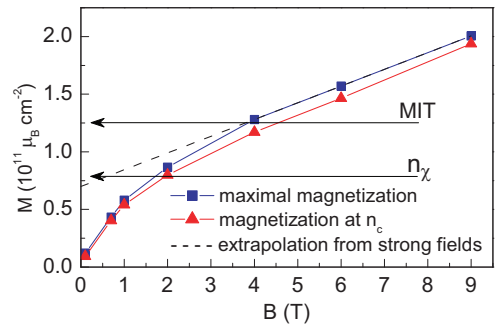}}
\caption{\label{fig1} Data of Prus {\it et al}~\cite{prus03}. The onset of strong localization in $B=0$ in their sample and the critical point $n_\chi$ are indicated by arrows. The data points obtained at low magnetic fields lie in the insulating regime, where the physics of local moments dominates.}
\end{figure}

In Fig.~1, we plot the main data of Prus {\it et al}. One can easily see that the low-$B$ data points lie in the insulating regime, where the physics of local moments dominates \cite{dolgopolov02}. (The $y$-axis is converted into electron density $n$ in accordance with the relation for the maximum magnetization $M=\mu_Bn$ \cite{prus03,reznikov04}, and the field $B$ corresponds to the onset of full spin polarization.) Based on the data obtained in the regime of {\em strong localization}, one cannot make judgments concerning the properties of a clean electron system. To this end, the attempt by Reznikov and Sivan to extend the analysis \cite{prus03} to the clean limit is not justified.

The authors of the Comment are confused about the location of the
point where the full spin polarization sets in and contradict
themselves contrasting ``maximal'' and ``full'' magnetizations.
Evidently, the onset of full spin polarization occurs at the point
where $M(n)$ reaches a maximum ({\it i.e.}, $dM/dn=0$ for nearly
anti-symmetric jumps in $dM/dn$ \cite{shashkin04}), reflecting the
beginning of the filling of the second spin subband. The method of
extracting spin susceptibility, described in the Comment, which
involves integration of $dM/dn$ from 0 to $n_m$ and use of the
formula $\mu_Bn_m(B)/B$, has nothing to do with the procedure we have
used in our paper \cite{shashkin04}.

Concerning the possible influence of the diamagnetic shift,
we have established experimentally that it is negligible in our case, as
follows from the concurrence of results obtained by different methods
including the magnetocapacitance method \cite{shashkin04}. The
attempt by Reznikov and Sivan to introduce their theoretical
estimates of the diamagnetic shift into the experimental data makes
little sense.

\end{document}